# Ultra-broadband photodetection of Weyl semimetal TaAs up to infrared 10 μm range at room temperature

Shumeng Chi[1], Zhilin Li[2], Ying Xie[1], Yongguang Zhao[3], Zeyan Wang[1], Lei Li[3], Haohai Yu[1]*, Gang Wang[2]**, Hongming Weng[2,4]***, Huaijin Zhang[1]****  Jiyang Wang[1]

[1]State Key Laboratory of Crystal Materials and Institute of Crystal Materials, Shandong University, Jinan 250100, China.

[2]Beijing National Laboratory for Condensed Matter Physics, Institute of Physics, Chinese Academy of Sciences, Beijing 100190, China.

[3]Jiangsu Key Laboratory of Advanced Laser Materials and Devices, School of Physics and Electronic Engineering, Jiangsu Normal University, Xuzhou 221116, China

[4]Collaborative Innovation Center of Quantum Matter, Beijing, China

* E-mail: haohaiyu@sdu.edu.cn

** E-mail: gangwang@iphy.ac.cn

***E-mail: hmweng@iphy.ac.cn

****E-mail: huaijinzhang@sdu.edu.cn

**Abstract**

Photodetectors with broadband optical response have promising applications in many advanced optoelectronic and photonic devices. Especially, those with the detection range up to mid-infrared at room temperature are very challenging and highly desired. Recently, Weyl semimetal has been discovered and proposed to be favorable for photodetection since in general it breaks Lorentz invariance to have tilted chiral Weyl cones around Fermi level, which leads to chirality dependent photocurrents at arbitrarily long wavelength. Furthermore, the linear dispersion bands in Weyl cones result in very high carrier mobility and much reduced thermal carrier compared with the parabolic ones in narrow-gap semiconductors. Here, we report that the Weyl semimetal TaAs based photodetector can operate at room temperature with spectral range from blue (438.5 nm) to mid-infrared (10.29 μm) light wavelengths and the responsibility and detectivity is more than 78 μA W$^{-1}$ and $1.88 \times 10^7$ Jones, respectively. This is the first photodetector made by a Weyl semimetal and shows its promising in room-temperature mid-infrared photodetection.

**Introduction**

The photodetectors which can convert the light to electric signals have attracted many interests since the discovery of photoelectric effects in 1890s[1-6]. Those with broadband optical response, ranging from ultraviolet, through visible to infrared and even terahertz, are of fundamental significance for many applications, such as imaging, sensing systems, environmental monitoring, optical communications and analytical applications[7-13]. According to the photodetection theory[14], the semiconductors with band gap smaller than the photon energy of incident light can have their electrons excited from the valance bands to conduction bands to achieve the photoresponse, which boosted the development of photodetectors based on narrow-gap semiconductors such as HgCdTe, PbS and PbSe for the detection of broadband lights, especially the mid-infrared photons[15,16]. However, for semiconductors with band gap comparable with or smaller than the mid-infrared light, thermal carriers at room-temperature would generate large dark currents overwhelming the photo-generated current, which limits the operation of photodetectors fabricated with narrow-gap semiconductors in high cost cryogenic temperature. This presents a big challenge for the photodetection in the mid-infrared range at room-temperature.

Recently, graphene emerges as a promising candidate for ultra-broadband photodetectors due to its gapless band structure and quite large carrier mobility ($10^4$ $cm^2$ $V^{-1}$ $s^{-1}$ at room temperature). But the low absorption efficiency, ~2.3% per layer, constrains its efficiency in converting light to the electric signal and the generation of detectable carriers at room temperature[17-19]. So far, the broadband graphene-based photodetectors with the detecting range from 532 nm to 10.31 μm have been realized only under the temperature of 10 K, but the photodetecting signals disappeared when the temperature is higher than 150 K[8]. Hence, several extensive efforts to improve the light absorptance and the photoresponse performance of graphene have been made, including the creation of a finite band gap and the introduction of various defect states but at the sacrifice of the broadband characteristics.

Weyl semimetal, a recently discovered new quantum state of metal[20-25], has been proposed to be of great potential value for new-generation photoelectric and electric devices since it hosts massless chiral Weyl fermions with linear relationship between energy and momentum. Due to this unusual electronic structure, three dimensional Weyl semimetal has been expected to have several distinct advantages for a photodetector when compared with narrow-gap semiconductor and graphene. Weyl semimetals have broadband photon absorption with extremely large absorption coefficients, chirality dependent photocurrent,[26] extremely high carrier mobility (~$10^4$ $cm^2$ $V^{-1}$ $s^{-1}$ comparable with graphene) and largely reduced thermal carrier under room temperature. What's more, the advantages of Weyl semimetals can be maintained in three dimensions, different from graphene whose band gap varies and photoresponse properties deteriorate with the layers. Inspired by these, we investigated a photodetector made by Weyl semimetal TaAs in a broadband regime (from visible 438.5 nm to the mid-infrared 10.29 μm) under room temperature. The responsibility and detectivity at the ambient temperature within such ultra-broadband regime is more than 78 μA $W^{-1}$ and $1.88 \times 10^7$ Jones, respectively. This reveals that Weyl semimetals can be very promising materials for broadband photodetectors as advanced photoelectric and electric devices.

**Results and Discussion**

The absorption spectrum of the Weyl semimetal TaAs in the wavelength range from 400 nm to 14 μm was characterized at room temperature as shown in Figure 1a, which is obtained by spectral diffuse reflectance of TaAs powder and achieved through the Kubelka–Munk method[27]. TaAs has dispersive absorption with varying incident photon energy and the absorptance decreases with the wavelength, which is consistent with the previous results on its optical spectroscopy and the analysis about its bulk electronic states. This indicates that with the decrease of incident photon energy, the electron transition process from the valence to conduction bands occurs closer to the Weyl nodes and reduces the absorption[28]. Based on the calibration with a 1030 nm laser[29], the absorption coefficient 2α varied from $48 \times 10^4$ cm$^{-1}$ to $6.35 \times 10^7$ cm$^{-1}$ with the penetration depth $\frac{1}{\alpha}$ from 42 nm to 0.315 nm in case the wavelength ranging from 400 nm to 14 µm. This means that the incident photons can be efficiently absorbed and is favorable to the broadband photodetection .

A TaAs single crystal with a rectangular section (2 mm × 2 mm) and the thickness of 1.1 mm along the (001) direction was employed for the characterization of its photodetecting performance. The incident light was normally projected onto the (001) surface as shown in Figure 1b. TaAs absorbs the photons and gives rise to the transitions of electrons and generates carriers, which leads to a photocurrent gain, the discrepancy between current under light illumination and in dark. A pair of Cu electrodes were utilized in the contact with the TaAs sample. Ohmic contact between the photocurrent gain sample and electrodes is necessary for efficient transfer of electrons from the gain medium to electrodes during photodetection. The dark current (*I*) without light illumination was measured with the varying applied voltage (*V*) from 0 to 100 µV. The *I-V* curve is presented in Figure 1c and its linearity illustrates that there is no Schottky barrier at the contacting interfaces between the TaAs and Cu electrodes[30], and the work function of TaAs is close to that of Cu with the value of 4.65 eV.

The photodetection performance of TaAs was investigated at room temperature in the wavelength range from the visible 438.5 nm to mid-infrared 10.29 µm with a 2450 Sourcemeter (also used as a bias source). The illumination sources at 438.5 nm, 963.6 nm, 2.02 µm, 2.82 µm and 10.29 µm are continuous-wave lasers and those at 3.02 µm,

4.07 μm, 5.06 μm, and 5.78 μm are pulsed lasers with the repetition rate of 1 MHz. The light on-off photocurrents at different incident light power are shown in Figure 2a-e and Figure 3a-d with the same applied voltage of 100 μV at the time interval of 30 s. Under the illumination with continues-wave lasers, the photocurrent can be observed to linearly augment with the incident light power in Figure 2f-j. The maximum photocurrent is 10.48 μA at the wavelength of 438.5 nm under the incident power of 70 mW. Under the same illumination power, the photocurrent shows the minimum at the wavelength of 2.82 μm and above 2.82 μm, the photocurrent increases with the light wavelength up to 5.78 μm. For the photonresponse at 10.29 μm, the maximum photocurrent is about 2.5 μA under the incident light power of 66 mW. The linear relationship between the photocurrent and incident power is different from that of the graphene photodetector with an introduced midgap defect band, which implies that the photocurrent gain in TaAs is simply generated from the transition from the valance to conduction bands. In order to characterize the photodetection at other wavelengths, pulsed lasers at 3.02 μm, 4.07 μm, 5.06 μm and 5.78 μm were employed. Under the same incident laser of 15 mW, we found that the photocurrent increased with the wavelength and the maximum photocurrent is 1.5 μA at 5.78μm.

According to the mechanism of photoelectric effect, the photocurrent can be described as[31]

$$I_{ph} = AVqu\Delta n \tag{1}$$
$$\propto \frac{\eta P_{inc}}{h\nu}$$

where $A$ is the cross section area of the active layer; $V$ is the applied bias; $q$ is the unit electron charge ($1.6 \times 10^{-19}$ Coulombs); $\mu$ is the carrier mobility; $\Delta n$ is the photon-generated carriers density; $P_{inc}$ is the incident optical power; $h\nu$ is the photon energy; $\eta$ is the quantum efficiency (i.e., the number of carriers generated per photon). The absorptance of Weyl semimetal TaAs shows a downward trend with the decrease of incident photon energy, while the amount of incident photons increases under the same light power in the meantime. We think that the photon-generated carrier density depending on the wavelength is a result of several effects, such as the quantum efficiency, thermal effects and relaxation time of the electrons in conduction band. Therefore, the photocurrent was observed to attain a minimum value at the wavelength of 2.82 μm and to show a non-monotonic trend over the wavelength range from 438.5 nm to 10.29 μm.

To evaluate the ability of converting the light signals into electrical signals and detecting signal from noise at a certain wavelength, responsivity ($R_\lambda$) and detectivity ($D^*$) are commonly used as figures of merits. $R_\lambda$ and $D^*$ are defined as[32-34]

$$R_\lambda = \frac{I_{ph}}{P_{Light}} \tag{2}$$

$$D^* = \frac{R_\lambda}{\sqrt{2qI_{dark}}} = \frac{I_{ph}}{P_{Light}\sqrt{2qI_{dark}}} \tag{3}$$

Where $P_{Light}$ is the illumination light power and $I_{dark}$ is the current in dark. $R_\lambda$ has the unit of μA W$^{-1}$ and $D^*$ has the unit of Jones.

According to the equations (2) and (3), $R_\lambda$ and $D^*$ at different wavelengths are presented in Figure 4a-b under the same illumination power of 10 mW. Consistent with the variation of photocurrent, both the responsivity and detectivity reach the smallest values of 78 µA W$^{-1}$ and 1.88×10$^7$ Jones at the wavelength of 2.82 µm. Above 2.82 µm, they exhibit a slight upward trend with increasing the wavelength up to 5.78 µm. The highest responsivity was measured to be 179 µA W$^{-1}$ at the wavelength of 438.5 nm, corresponding to $D^*$ of 4.5×10$^7$ Jones. For the photodetecting performance at 10.29 µm, $R_\lambda$ and $D^*$ were measured to be 88 µA W$^{-1}$ and 2.34×10$^7$ Jones, respectively. To the best of our knowledge, this is the first photodetector based on Weyl semimetal and has the broadest range at room temperature. It is noted that the responsivity and detectivity of TaAs at 3.02 µm are 86 µA W$^{-1}$ and 2.13×10$^7$ Jones by using a pulsed laser, which is obviously higher than those at 2.82 µm measured with a continues-wave laser. Associated with the comparable photon energy 0.44 eV (at 2.82 µm) and 0.41 eV (at 3.02 µm), the discrepancy of $R_\lambda$ and $D^*$ at the two cases may result from less thermal generation in the process of photocurrent gain under illumination of pulsed lasers, which also indicates that the photodetection performance of TaAs could be further improved under lower temperature. In addition, the responsivities of the Weyl semimetal TaAs at different wavelengths all show a downward trend with increasing incident power (see Supporting Information), indicating that the photocurrent is a result of linear processes such as thermal generation or trap-mediated absorption.

Figure 5 presents the operation spectrum and operation temperature of photodetectors based on Weyl semimetal TaAs, pure monolayer graphene and some commonly used semiconductor materials, showing that TaAs has the broadest band photoresponse (from blue to mid-infrared) under room temperature, while for other materials, broadband photodetection only can be achieved at extremely low temperature. The results presented above raise the question of why TaAs can realize photodetection up to mid-infrared or even longer wavelength under room temperature. Chan et al. had discussed several important points in Ref. 26, such as the breaking of both inversion and Lorentz symmetries with tilted Weyl cones, the tuning of chemical potential and incident light direction. In addition to these, the linear dispersion around Weyl cones leads to high carrier mobility and quadratic, instead of square root in semiconductors, energy dependence of density of states (DOS). The high mobility enhances the efficiency of photocurrent and the reduced DOS around Weyl nodes suppresses the thermal carrier. The absorption coefficient is also important to material intended for infrared photodetectors[35]. TaAs has absorption coefficient of $4.8 \times 10^5$ cm$^{-1}$ with the penetration depth of 42 nm at light wavelength of 1030 nm[29], close to $6.8 \times 10^5$ cm$^{-1}$ of graphene[17], which is desired for broadband and high detectivity photodetectors[26].

In summary, we have experimentally demonstrated the ultra-broadband photoresponse of Weyl semimetal TaAs under room temperature. Compared with graphene and semiconductors, TaAs definitely has the advantages in the broadest detection region and the ambient working temperature based on its massless chiral Weyl nodes and fairly large absorptance over a wide spectral range. This work opens up exciting opportunities for the application of TaAs in the photondetection.

**Methods**

**Light sources.** In this work, continues-wave lasers with wavelengths of 438.5 nm,

963.65nm were generated from different continues-wave fiber-coupled laser diodes. 2.018 μm wavelength light was obtained using a Tm: YGG crystal under the pump of a 795nm fiber-coupled laser diode. 2.82 μm wavelength light was obtained using a Er: YGG crystal under the pump of a 976nm fiber-coupled laser diode. 10.29 μm was generated using a $CO_2$ laser. Pulsed lasers of 3.02 μm, 4.07 μm, 5.06 μm and 5.78 μm were provided by an optical parametric amplification (OPA) with a pulse having width of 220 fs, and repetition rate of 1 MHz.

**TaAs growth and preparations.** Large-size and high quality Single crystals of TaAs were grown through the chemical vapor transport (CVT) method[36]. First, Tantalum foil (99.99%), arsenic pieces (99.995%) and the transport agent iodine (99.99%) were put into a silica ampule filled of argon (mole ratio Ta:As:$I_2$ =1:1:0.05). After evacuating and sealing, the mixture in the silica ampule was heated to 1000 ℃ over 72 h and TaAs polycrystalline was produced. Then the ampoule was kept in a temperature field (the hot and cold ends were held at 1020 °C and 980 ℃) for two weeks and finally cooled down to room temperature by shut down the power. The (001) surface of the as-grown 1.1 mm thick TaAs single crystal is used for the photoresponse measurements.

**Acknowledgements**

The authors would like to thank Prof. X.L. Chen and Prof. L.W. Guo of Institute of Physics, CAS for the fruitful discussions. This work was supported by the National Natural Science Foundation of China (Grant Nos. 51422205, 51322211, 51572291, 11674369 and 11422428), the Strategic Priority Research Program (B) of the Chinese

Academy of Sciences (Grant No. XDB07020100), Natural Science Foundation for Distinguished Young Scholars of Shandong Province (JQ201415), Taishan Scholar Foundation of Shandong Province, China. H.W. is also supported by the National Key Research and Development Program of China (Grant No. 2016YFA0300600).

**Figures**

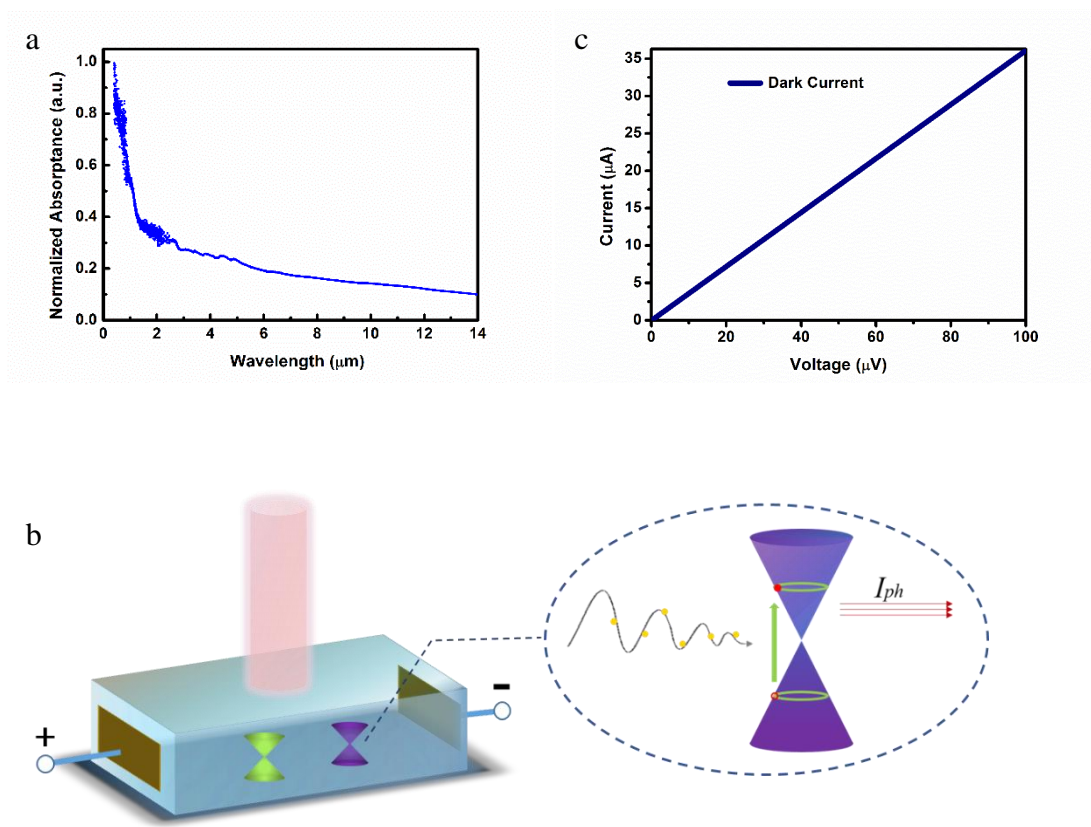

**Figure 1** (a) The normalized absorptance of TaAs as a function of wavelength. The absorptance declines with the increase of wavelength and tends to be gently in the mid-infrared regime. (b) Schematic of the experimental setup. The focused laser beams were shed into TaAs in normal incidence condition. Continues-wave lasers and pulsed lasers were employed as light sources. The laser spots were focused to ~1mm and ~2 mm for continues-wave lasers and pulsed lasers, respectively. (c) The I-V curve of the device. Measurements were conducted by applying bias voltage from 0 V to 100 μV. The curve is linear, which demonstrates that there is an efficient electronic transport path between the electrodes and TaAs sample.

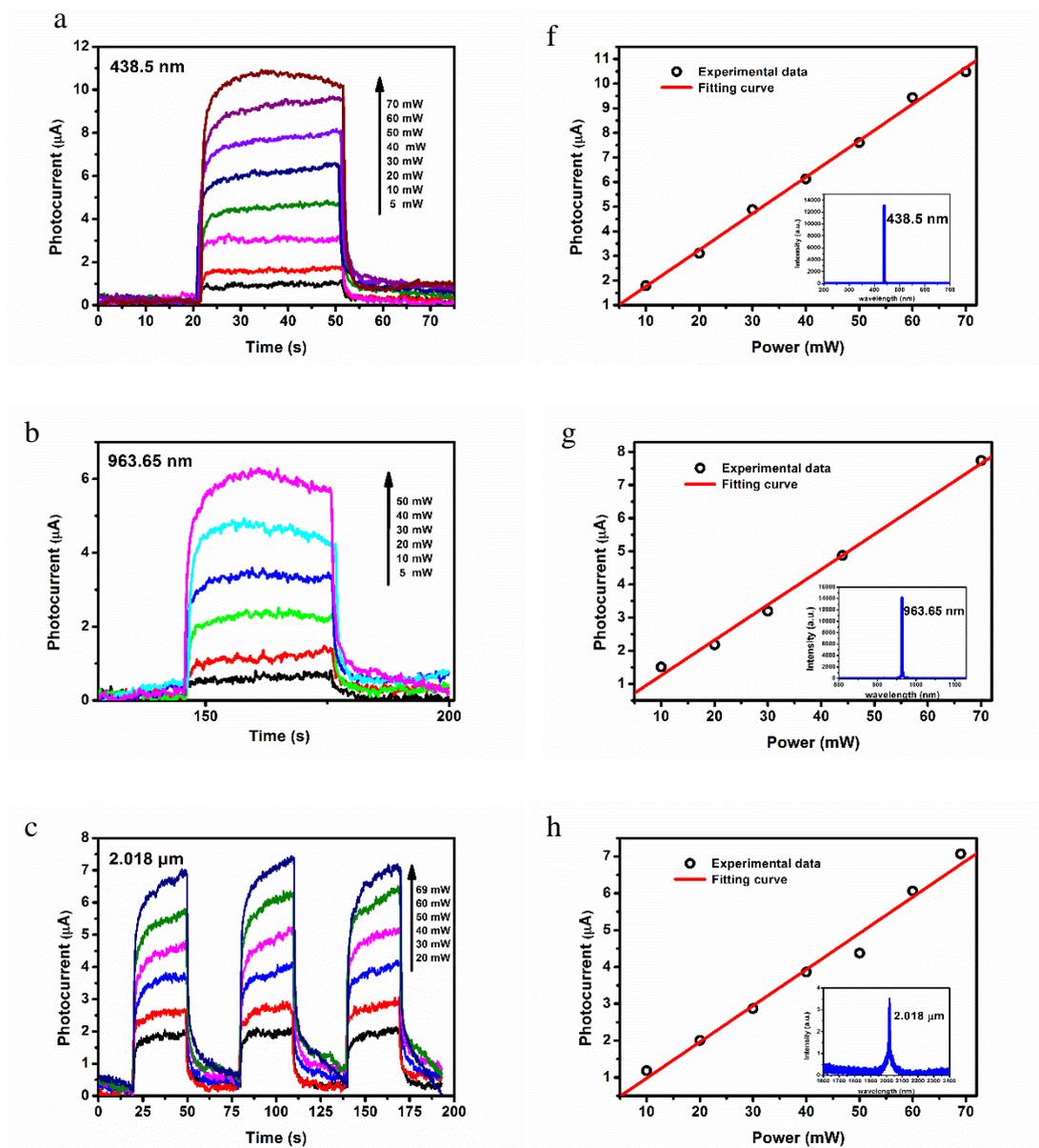

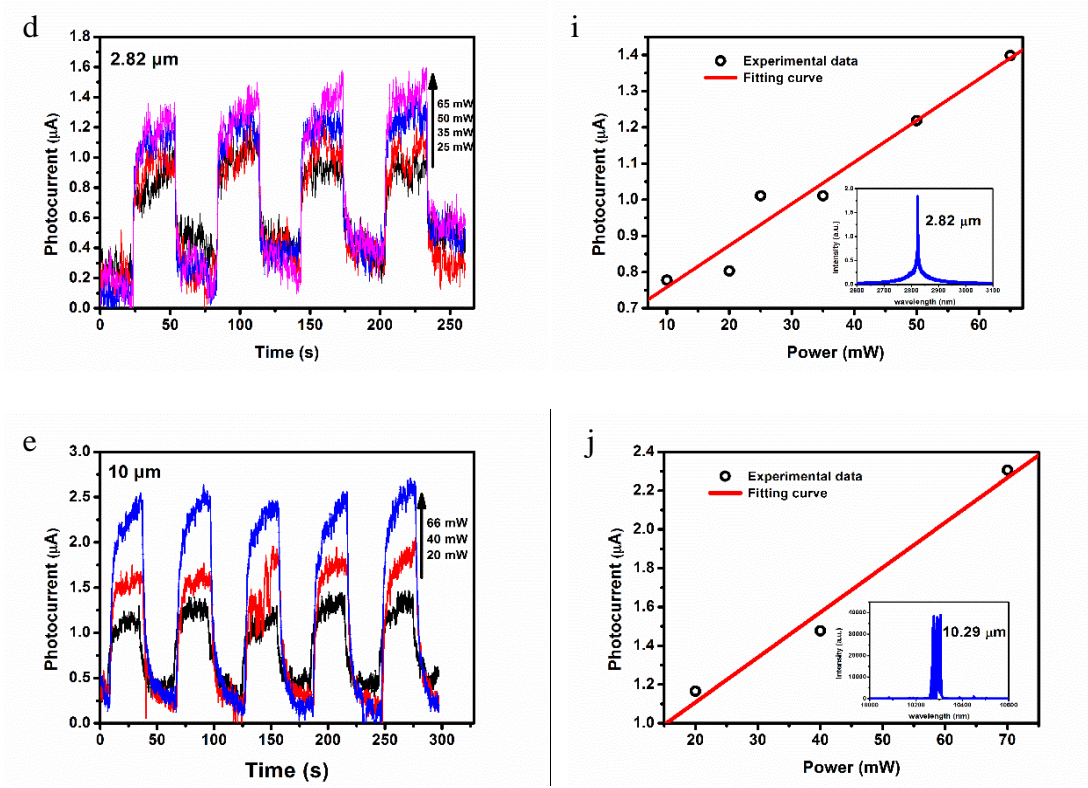

**Figure 2** Photoresponse features of TaAs under excitation of continues-wave laser at a fixed bias voltage of 100 μV. Time-dependent photocurrent of TaAs at wavelengths of (a) 438.5 nm, (b) 963.65 nm, (c) 2.018 μm, (d) 2.82 μm, (e) 10.29 μm with periods of on-off operation at the time interval of 30 s. Photocurrent versus incident power at (f) 438.5 nm, (g) 963.65 nm, (h) 2.018 μm, (i) 2.82 μm, (j) 10.29 μm. The inset in (f)-(j) are spectrums of the excitation wavelengths.

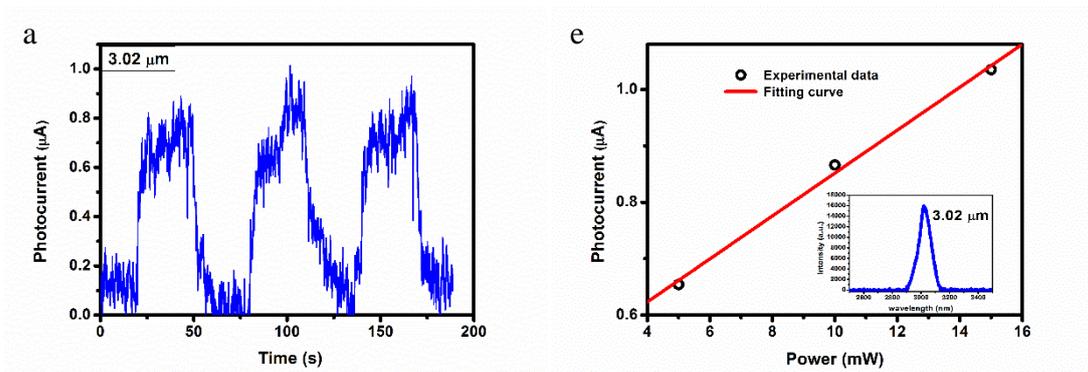

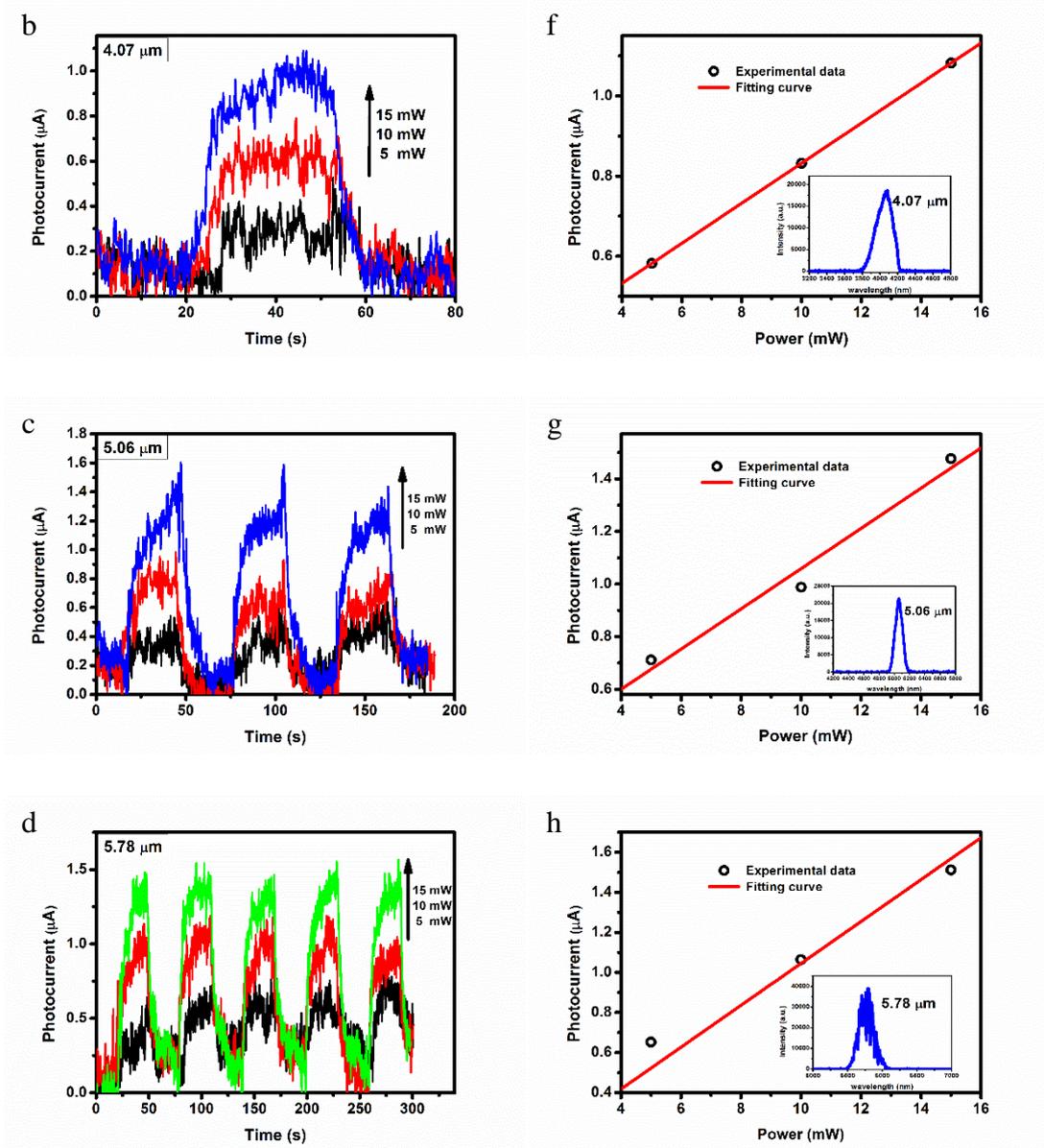

**Figure 3** Photoresponse features of TaAs sample under excitation of pulsed laser at a fixed voltage bias of 100 μV. Time-dependent photocurrent of TaAs at wavelengths of (a) 3.02 μm, (b) 4.07 μm, (c) 5.06 μm and (d) 5.78 μm with periods of on-off operation at the time interval of 30 s. Photocurrent as a function of incident power at (e) 3.02 μm, (f) 4.07 μm, (g) 5.06 μm and (h) 5.78 μm, where both axes are in the logarithmic scale. The inset in (e)-(h) are spectrums of the excitation wavelengths.

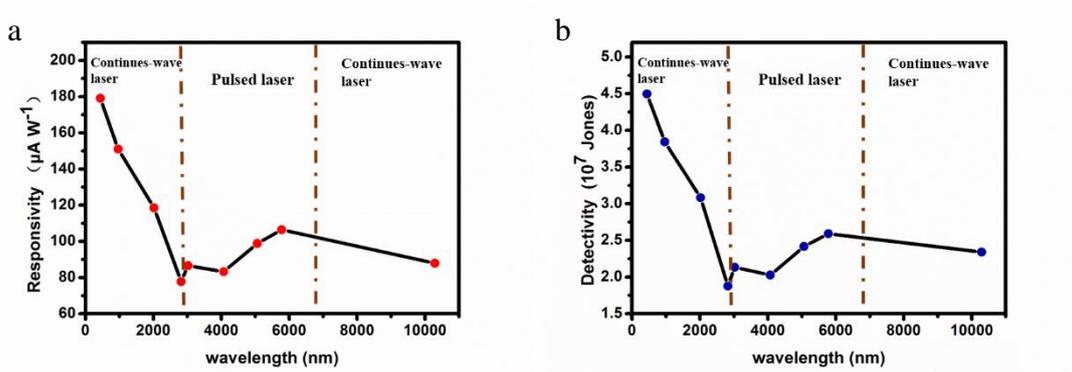

**Figure 4**  The (a) responsivity $R_\lambda$ and (b) detectivity $D^*$ of TaAs as a function of excitation wavelengths under the same pump power (10 mW). Under continues-wave lasers, $R_\lambda$ and $D^*$ decrease with increasing the wavelength, while exhibit a slight upward trend under pulsed lasers pumping, indicating a complex process of carries generation and transfer.

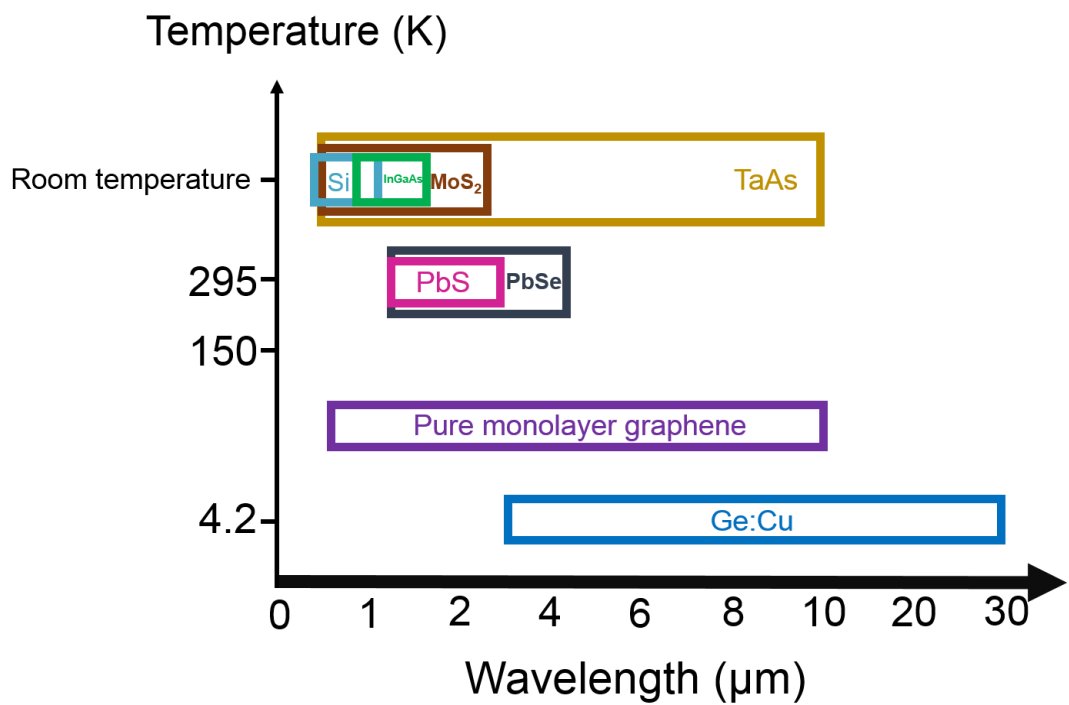

**Figure 5**  Operation spectrum and operation temperature of photodetectors based on different materials.

**References**

1. Konstantatos G, Sargent EH. Nanostructured materials for photon detection. *Nature nanotechnology* 2010, **5**(6): 391-400.

2. Tang L, Kocabas SE, Latif S, Okyay AK, Ly-Gagnon D-S, Saraswat KC*, et al.* Nanometre-scale germanium photodetector enhanced by a near-infrared dipole antenna. *Nature Photonics* 2008, **2**(4): 226-229.

3. Hu L, Yan J, Liao M, Xiang H, Gong X, Zhang L*, et al.* An Optimized Ultraviolet-A Light Photodetector with Wide-Range Photoresponse Based on ZnS/ZnO Biaxial Nanobelt. *Advanced Materials* 2012, **24**(17): 2305-2309.

4. O'Brien GA, Quinn AJ, Tanner DA, Redmond G. A single polymer nanowire photodetector. *Advanced materials* 2006, **18**(18): 2379-2383.

5. Pospischil A, Humer M, Furchi MM, Bachmann D, Guider R, Fromherz T*, et al.* CMOS-compatible graphene photodetector covering all optical communication bands. *Nature Photonics* 2013, **7**(11): 892-896.

6. Jeong I-S, Kim JH, Im S. Ultraviolet-enhanced photodiode employing n-ZnO/p-Si structure. *Applied physics letters* 2003, **83**(14): 2946-2948.

7. Liu C-H, Chang Y-C, Norris TB, Zhong Z. Graphene photodetectors with ultra-broadband and high responsivity at room temperature. *Nature nanotechnology* 2014, **9**(4): 273-278.

8. Zhang Y, Liu T, Meng B, Li X, Liang G, Hu X*, et al.* Broadband high photoresponse from pure monolayer graphene photodetector. *Nature communications* 2013, **4:** 1811.

9. Gassenq A, Gencarelli F, Van Campenhout J, Shimura Y, Loo R, Narcy G*, et al.* GeSn/Ge heterostructure short-wave infrared photodetectors on silicon. *Optics express* 2012, **20**(25): 27297-27303.

10. Tsai D-S, Lin C-A, Lien W-C, Chang H-C, Wang Y-L, He J-H. Ultra-high-responsivity broadband detection of Si metal–semiconductor–metal schottky photodetectors improved by ZnO nanorod arrays. *ACS nano* 2011, **5**(10): 7748-7753.

11. Adhikary S, Aytac Y, Meesala S, Wolde S, Unil Perera A, Chakrabarti S. A

multicolor, broadband (5–20 μm), quaternary-capped InAs/GaAs quantum dot infrared photodetector. *Applied Physics Letters* 2012, **101**(26)**:** 261114.

12. Matthew Menke S, Pandey R, Holmes RJ. Tandem organic photodetectors with tunable, broadband response. *Applied Physics Letters* 2012, **101**(22)**:** 223301.

13. Nanot S, Cummings AW, Pint CL, Ikeuchi A, Akiho T, Sueoka K*, et al.* Broadband, polarization-sensitive photodetector based on optically-thick films of macroscopically long, dense, and aligned carbon nanotubes. *Scientific reports* 2013, **3:** 1335.

14. Sun Z, Chang H. Graphene and graphene-like two-dimensional materials in photodetection: mechanisms and methodology. *Acs Nano* 2014, **8**(5)**:** 4133-4156.

15. Popa A, Lisca M, Stancu V, Buda M, Pentia E, Botila T. Crystallite size effect in PbS thin films grown on glass substrates by chemical bath deposition. *Journal of Optoelectronics and Advanced Materials* 2006, **8**(1)**:** 43.

16. Norton P. HgCdTe infrared detectors. *Optoelectronics review* 2002(3)**:** 159-174.

17. Tsai D-S, Liu K-K, Lien D-H, Tsai M-L, Kang C-F, Lin C-A*, et al.* Few-layer MoS2 with high broadband photogain and fast optical switching for use in harsh environments. *Acs Nano* 2013, **7**(5)**:** 3905-3911.

18. Schwierz F. Graphene transistors. *Nature nanotechnology* 2010, **5**(7)**:** 487-496.

19. Dean CR, Young AF, Meric I, Lee C, Wang L, Sorgenfrei S*, et al.* Boron nitride substrates for high-quality graphene electronics. *Nature nanotechnology* 2010, **5**(10)**:** 722-726.

20. Weng H, Yu R, Hu X, Dai X, Fang Z. Quantum anomalous Hall effect and related topological electronic states. *Advances in Physics* 2015, **64**(3)**:** 227-282.

21. Weng H, Fang C, Fang Z, Bernevig BA, Dai X. Weyl Semimetal Phase in Noncentrosymmetric Transition-Metal Monophosphides. *Physical Review X* 2015, **5**(1)**:** 011029.

22. Lv B, Xu N, Weng H, Ma J, Richard P, Huang X*, et al.* Observation of Weyl nodes in TaAs. *Nature Physics* 2015.

23. Lv BQ, Weng HM, Fu BB, Wang XP, Miao H, Ma J, *et al.* Experimental Discovery of Weyl Semimetal TaAs. *Physical Review X* 2015, **5**(3)**:** 031013.

24. Xu S-Y, Belopolski I, Alidoust N, Neupane M, Bian G, Zhang C, *et al.* Discovery of a Weyl fermion semimetal and topological Fermi arcs. *Science* 2015, **349**(6248)**:** 613-617.

25. Huang S-M, Xu S-Y, Belopolski I, Lee C-C, Chang G, Wang B, *et al.* A Weyl Fermion semimetal with surface Fermi arcs in the transition metal monopnictide TaAs class. *Nature Communications* 2015, **6:** 7373.

26. Chan C-K, Lindner NH, Refael G, Lee PA. Photocurrents in Weyl semimetals. *arXiv preprint arXiv:160707839* 2016.

27. Kudo A, Ueda K, Kato H, Mikami I. Photocatalytic O2 evolution under visible light irradiation on BiVO4 in aqueous AgNO3 solution. *Catalysis Letters* 1998, **53**(3)**:** 229-230.

28. Xu B, Dai YM, Zhao LX, Wang K, Yang R, Zhang W, *et al.* Optical spectroscopy of the Weyl semimetal TaAs. *Physical Review B* 2016, **93**(12)**:** 121110.

29. Chi S, Li Z, Yu H, Wang G, Wang S, Zhang H, *et al.* Symmetrical broken and nonlinear response of Weyl semimetal TaAs influenced by the topological surface states and Weyl nodes. *Annalen der Physik* 2017, **529**(4)**:** 1600359-n/a.

30. Heinze S, Tersoff J, Martel R, Derycke V, Appenzeller J, Avouris P. Carbon Nanotubes as Schottky Barrier Transistors. *Physical Review Letters* 2002, **89**(10)**:** 106801.

31. Sze SM, Ng KK. *Physics of semiconductor devices*. John wiley & sons, 2006.

32. Gong X, Tong M, Xia Y, Cai W, Moon JS, Cao Y, *et al.* High-detectivity polymer photodetectors with spectral response from 300 nm to 1450 nm. *Science* 2009, **325**(5948)**:** 1665-1667.

33. Choi W, Cho MY, Konar A, Lee JH, Cha GB, Hong SC, *et al.* High‐detectivity multilayer MoS2 phototransistors with spectral response from ultraviolet to infrared. *Advanced Materials* 2012, **24**(43)**:** 5832-5836.

34. Dou L, Yang YM, You J, Hong Z, Chang W-H, Li G, *et al.* Solution-processed hybrid perovskite photodetectors with high detectivity. *Nature* *communications* 2014, **5**.

35. Martyniuk P, Rogalski A. HOT infrared photodetectors. *Opto-Electronics Review* 2013, **21**(2)**:** 239-257.

36. Li Z, Chen H, Jin S, Gan D, Wang W, Guo L*, et al.* Weyl Semimetal TaAs: Crystal Growth, Morphology, and Thermodynamics. *Crystal Growth & Design* 2016, **16**(3)**:** 1172-1175.